\documentclass[aps,prd,superscriptaddress,showpacs,preprint]{revtex4}
\usepackage{graphicx, bm}
\begin{document}
\draft
\title{Stellar energy loss rates in the pair-annihilation process beyond the standard model}

\author{ M. A. Hern\'andez-Ru\'{\i}z\footnote{mahernan@uaz.edu.mx}}
\affiliation{\small Unidad Acad\'emica de Ciencias Qu\'{\i}micas, Universidad Aut\'onoma de Zacatecas\\
         Apartado Postal C-585, 98060 Zacatecas, M\'exico.\\}

\author{A. Guti\'errez-Rodr\'{\i}guez\footnote{alexgu@fisica.uaz.edu.mx}}
\affiliation{\small Facultad de F\'{\i}sica, Universidad Aut\'onoma de Zacatecas\\
             Apartado Postal C-580, 98060 Zacatecas, M\'exico.\\}

\author{A. Gonz\'alez-S\'anchez\footnote{agonzalez@fisica.uaz.edu.mx}}
\affiliation{\small Facultad de F\'{\i}sica, Universidad Aut\'onoma de Zacatecas\\
             Apartado Postal C-580, 98060 Zacatecas, M\'exico.\\}
             
\affiliation{\small LERMA, CNRS UMR 8112, PSL Research University, Observatoire de Paris, 61 Avenue de l’Observatoire,
             75014, Paris, France.}             

\date{\today}

\begin{abstract}

We calculate the stellar energy loss due to neutrino-pair production in $e^+e^-$ annihilation
in the context of a 331 model, a left-right symmetric model and a simplest little Higgs model
in a way that can be used in supernova calculations. We also present some simple estimates which
show that such process can act as an efficient energy loss mechanism in the shocked supernova core.
We find that the stellar energy loss is almost independent on the parameters of the models in the
allowed range for these parameters. This work complements other studies on the stellar energy loss
rate in $e^+e^-$ annihilation.

\end{abstract}


\vspace{5mm}

\maketitle


\section{Introduction}

Gamow \cite{Gamow,Gamow1} and Pontecorvo \cite{Pontecorvo} were the first to recognize the important role played by neutrinos
in the evolution of stars. The neutrino emission processes may affect the properties of matter at high temperatures, and hence
affect stellar evolution.

On the other hand, when a massive star collapses in an explosion of a supernova, almost $99\%$ of the energy released
comes out in the form of neutrinos, with only $1-2\%$ coming out as light. Many of these neutrinos have energies of the
order of $10-30\hspace{0.8mm}MeV$. This results in much more neutrinos being produced in a few seconds that all those
released in the rest of the star life time. These neutrinos are produced in all flavors $(\nu_e, \nu_\mu, \nu_\tau)$ and
about the same number of particles  than antiparticles. Among the material ejected during the explosion there are heavy
elements that are important for the stellar evolution of galaxies, stars, planets and life. Other supernovas can create
neutron stars, remnants or even black holes depending on the mass of the star. In general, the neutrinos radiated by the
supernovas carry in their spectrum key information not only about the detailed nature of the supernova collapse but also
about properties of neutrinos, not yet explored in the laboratories \cite{Mohapatra}. This is one reason why it is important
to study the stellar energy loss rates due to neutrino pair production in annihilation $e^+e^-$.

Neutrino emission is known to play an important role in stellar evolution, especially in the late stages when the rate of
evolution is almost fully dependent on the energy loss via neutrinos. This refers to the stage of steady burning prior to the
implosion of the stellar core, to the process of catastrophic core-collapse, and to the cooling of the neutron star which
is formed.

The stellar energy loss rate due to neutrino emission receives contributions from both, weak nuclear reactions and purely leptonic
processes. However, for the large values of density and temperature which characterize the final stage of stellar
evolution, the latter are largely dominant, and are mainly produced by four possible interaction mechanisms \cite{Gilles,Dicus,Duane,Alam,Dicus1,Bruenn}:

\begin{eqnarray}
e^+ + e^- &\to& \nu+\bar \nu \hspace{3mm} \mbox{(pair annihilation)},\\
\gamma + e^\pm &\to&  e^\pm + \nu + \bar \nu \hspace{3mm} \mbox{($\nu$-photoproduction)},\\
\gamma^* &\to&  \nu + \bar \nu \hspace{3mm} \mbox{(plasmon decay)},\\
e^\pm + Z &\to& e^\pm + Z + \nu + \bar \nu \hspace{3mm} \mbox{(bremsstrahlung on nuclei)}.
\end{eqnarray}

Actually  these processes are the dominant cause of the energy loss rate in different regions in a density-temperature plane. For
very large core temperature, $T\gtrsim 10^{9}$ $^oK$, and not excessively high values of density, pair annihilations are most
efficient, while $\nu$ photoproduction gives the leading contribution for $10^{8}$ $^oK$ $ \lesssim T\lesssim 10^{9}$ $^oK$
and relatively low density, $\rho \lesssim 10^5$ $g$ $cm^{-3}$. These are the typical ranges for very massive stars in their late
evolution \cite{Gilles,Dicus,Duane,Alam}.

The Standard Model (SM) \cite{SM,SM1,SM2} of the electroweak interactions is the starting point of all the extended gauge models. In other words, any
gauge group with physical sense must have as a subgroup the $SU(2)_L\times U(1)_Y$ group of the standard model. The purpose of the extended theories
is to explain some fundamental aspects which are not clarified in the frame of the SM. One of these aspects is the origin of parity violation at current energies. The left-right symmetric models (LRSM) based on the $SU(2)_R\times SU(2)_L\times U(1)_Y$ gauge group \cite{Pati,Mohapatra0,Mohapatra1,Senjanovic,Senjanovic1} give an answer to that problem, since they restore the parity symmetry at high energies and give their violations at low energies as a result of the breaking of gauge symmetry. Detailed discussions on the left-right symmetric model can be found in the literature \cite{Pati,Mohapatra0,Mohapatra1,Senjanovic,Senjanovic1,Mohapatra2}.

The $SU(3)_C \times SU(3)_L\times U(1)_X$ model \cite{Frampton,Pisano}, also called 331 model is one of the most simplest and attractive extensions
of the SM. In the literature \cite{Dias,Dias1,Diaz,Dias2,Dias3,Dias4,Dias5,Dias6,Ochoa,Cogollo,Qing} there are different versions of this model which are characterized by the parameter ${\ss}=\pm \sqrt{3}$ and ${\ss}=\pm \frac{1}{\sqrt{3}}$. The different models with different choices of ${\ss}$ have new particles with different electric charges. However, in general these models have the same characteristics, that is to say: 1) Unlike the SM that anomaly cancellation is fulfilled within each generation, the gauge anomaly is cancelled in the 331 model when considering all the generations. The number of generations $N$ must be a multiple of three. On the other hand, in order to ensure $QCD$ an asymptotic free theory, $N$ has to be smaller than six. So the number of generations $N$ is equal to three, which explains why the SM has three generations. 2) One of the three quark generations is different from the
other two, making sure that the anomaly is free, which leads to tree-level Flavour Changing Neutral Current (FCNC) through a new neutral gauge boson $Z'$
or the mixing $Z-Z'$. 3) Peccei-Quinn (PQ) symmetry \cite{Pecei} which can solve the strong CP problem is a natural result of gauge invariance in the 331
model \cite{Dias,Dias1}. 4) As a consequence of the extended gauge sector, the 331 model contains a much broader spectrum of particles than the SM: more
heavy quarks or leptons, more gauge bosons and more Higgs scalars. This may change the SM phenomenology significantly and lead to interesting signatures
at the current and future colliders such as the Large Hadron Collider (LHC) \cite{Aad,Aad1,Chatrchyan}, International Linear Collider (ILC) \cite{Abe,Aarons,Brau,Baer,Asner,Zerwas} and the Compact Linear Collider (CLIC) \cite{Accomando,Abramowicz,Dannheim}.

The existence of a heavy neutral ($Z'$) vector boson is a feature of many extensions of the standard model. In particular, one (or
more) additional $U(1)'$ gauge factor provides one of the simplest extensions of the SM. Additional $Z'$ gauge bosons appear in Grand
Unified Theories (GUTs) \cite{Robinett}, Superstring Theories \cite{Green}, Left-Right Symmetric Models (LRSM) \cite{Mohapatra0,G.Senjanovic,G.Senjanovic1}
331 model \cite{Frampton,Pisano} and in other models such as models of composite gauge bosons \cite{Baur}. In particular, it is possible to study some phenomenological features associates with this extra neutral gauge boson through little Higgs model. Many little Higgs models have been proposed in the literature; however the Littlest Higgs model (LH) proposed by N. Arkani-Hamed, {\it et al.} \cite{Arkani1,Arkani2}, provides one of the most economical implementations and forms the basis for most phenomenological analysis. The LH model \cite{Arkani1,Arkani2} has been proposed for solving the little hierarchy problem. In this scenario, the Higgs boson is regarded as a pseudo Nambu-Goldstone boson associated with a global symmetry at some higher scale. Though the symmetry is not exact, its breaking is specially arranged to cancel quadratically divergent corrections to the Higgs mass term at 1-loop level. This
is called the little Higgs mechanism. As a result, the scale of new physics can be as high as 10 $TeV$ without a fine-tuning on
the Higgs mass term. Among various little  Higgs models, the simplest little Higgs model (SLH) \cite{Kaplan,Schmaltz,Dias7} is attractive due to its relatively simple theory structure. Detailed discussions on the little Higgs models are reported in the literature \cite{Arkani1,Arkani2,Cheng,Cheng1,Low,Kaplan,Schmaltz,Dias7,Tao,Csaki,Hewett,Chen,Chen1,ILow}.

Our main objective in this work is to provide suitable expressions for the stellar energy loss rates of pair production of neutrinos
via the process $e^+ e^- \to \nu \bar \nu$ in the context of three models, a 331 Model (331M) \cite{Cogollo}, a Left-Right Symmetric Model (LRSM) \cite{Pati,Mohapatra0,Mohapatra1,Senjanovic,Senjanovic1,Polak,A.Gutierrez6} and the Simplest Little Higgs Model (SLHM) \cite{Dias7,Marandella}.
These will be expressed in a form which can be easily incorporated into realistic supernova models. These models have the interesting feature that they
are independent of the mass of the new additional $Z'$ heavy gauge boson, and only depend on the mixing angles $\theta$ and $\phi$ of the 331M and LRSM
and of the characteristic energy scale $f$ of the SLHM besides the SM parameters. For this reason, we chose these models to calculate
the stellar energy loss rates of neutrinos in supernova.

The neutrinos play a crucial role for the understanding of core-collapse supernova in terms of heating and cooling of supernova matter as well as for the  incompletely known supernova explosion mechanism \cite{Janka,Janka1,Naoki}. The long term neutrino signal of the deleptonizing/cooling nascent protoneutron star, which is to say after the supernova explosion has been launched, was reviewed in Refs. \cite{Huedepohl,Gava,Arnet,Fischer}. Both studies are milestones of consistent simulations of supernova explosions and represent standard works in the field of core-collapse supernova modeling. The associated long-term neutrino signal $\sim 10-30$ seconds is relevant for supernova neutrino detection, for recent insights see Ref. \cite{Wu}. Detailed analyses regarding the neutrino spectra formation and evolution including the neutrino-energy hierarchy can be found in the literature \cite{Raffelt2,Keil,Fischer1}.

Stellar energy loss rates data have been used to put constraints on the properties and interaction of light particles \cite{Dicus,Dicus1,Ellis,Bruenn}.
In addition, one of the most interesting possibilities to use stars as particle physics laboratories \cite{Raffelt} is to study the backreaction of the novel energy loss rates implied by the existence of new low-mass particles such as axions \cite{Payez,Fischer2}, or by non-standard neutrino properties such as magnetic moment and electric dipole moment \cite{Raffelt1,Kerimov,Alexander,Blinnikov}.

This paper is organized as follows: In Sect. II we present the calculation of the stellar energy loss rates of the process $e^+ e^- \to \nu \bar \nu$
for our three models. In Sect. III we give our results and conclusions.

\section{Stellar energy loss rates beyond the standard model}

\subsection{Stellar energy loss rates through $e^+ +e^- \rightarrow \nu +\bar{\nu}$ in a 331 model}

In the context of this model we obtain the energy loss rates through the pair-annihilation process

\begin{equation}
e^+(p_1)+e^-(p_2)\rightarrow \bar \nu (k_1, \lambda_1)+ {\nu}(k_2,\lambda_2),
\end{equation}

\noindent with $Z$ exchange, which is to say, in the limit of a four-fermion electroweak
interaction no electromagnetic radiative corrections. Here the $k_i$ and $p_i$ are the particle momenta and $\lambda$ is the
helicity of the neutrino.

The amplitude of transition for the process give in Eq. (5) is

\begin{equation}
{\cal M}=-\frac{g^2ab}{M^2_Z\cos^2\theta_W}\left[\bar{u}\left(k_2,\lambda_2\right)\gamma^\mu\frac{1}{2}\left( g_V^\nu- g_A^\nu
\gamma_5\right)v\left(k_1,\lambda_1\right)\right]\left[\bar{v}\left(p_1\right)\gamma_\mu\frac{1}{2}\left( g_V^e- g_A^e \gamma_5\right)u\left(p_2\right)\right],
\end{equation}

\noindent where the constant $a$ and $b$ depend only on the parameters of the 331M \cite{Cogollo}

\begin{equation} a=\cos{\theta}-\frac{\sin{\theta}}{\sqrt{3-4\sin^2\theta_W}} \hspace{5mm} \mbox{and}
\hspace{5mm} b=\cos{\theta}+\frac{(1-2\sin^2\theta_W)}{\sqrt{3-4\sin^2\theta_W}}\sin\theta,
\end{equation}

\noindent and where $\theta$ is the mixing angle between $Z-Z'$ of the SM and the 331M \cite{Cogollo}, $g$ is the coupling constant
and it related to the Fermi constant $G_F$ through the relation $G_F=\frac{\sqrt{2}g^2}{8M^2_W}=1.166 378 7(6)\times10^{-5}$ $GeV^{-2}$ \cite{Data2014},
with $M_W$ the mass of the charged $(W^\pm)$ vector boson, $u$ and $v$ are the usual Dirac spinors. We then write

\begin{equation}
\sum_s|{\cal M}|^2=
\frac{G_F^2}{2}a^2b^2{\cal N}^{\mu\nu}{\cal E}_{\mu\nu},
\end{equation}

\noindent where

\begin{eqnarray}
{\cal N}^{\mu\nu}&=&\frac{1}{4}Tr[( k\llap{/}_{2}+m_\nu)\gamma^\mu(g^\nu_V-g^\nu_A\gamma_5) ( k\llap{/}_{1}-m_\nu)\gamma^\nu(g^\nu_V-g^\nu_A\gamma_5)],\\
{\cal E}_{\mu\nu}&=&\frac{1}{4}Tr[( p\llap{/}_{2}+m_e)\gamma_\mu(g^e_V-g^e_A\gamma_5) ( p\llap{/}_{1}-m_e)\gamma_\nu(g^e_V-g^e_A\gamma_5)],
\end{eqnarray}

\noindent here $m_\nu$ and $m_e$ are the neutrino and electron mass, respectively.

We now evaluate the traces given in Eqs. (9) and (10) and the contraction of ${\cal N}^{\mu\nu} {\cal E}_{\mu\nu}$ gives

\begin{eqnarray}
N^{\mu\nu}E_{\mu\nu}=&16&\left\lbrace(g^e_V + g^e_A)^2(p_1\cdot k_1)(p_2\cdot k_2)+ (g^e_V - g^e_A)^2 (p_1\cdot k_2) (p_2\cdot k_1)\right.\nonumber\\
&+&\left.\left[(g^e_V)^2-(g^e_A)^2\right] m_e^2(k_1\cdot k_2)\right\rbrace,
\end{eqnarray}

\noindent where $g^e_V=-\frac{1}{2}+2\sin^2\theta_W$ and $g^e_A=-\frac{1}{2}$.

From Eqs. (8) and (11) the explicit form for the squared transition amplitude is

\begin{eqnarray}
\sum_s|{\cal M}|^2=&8G_F^2&a^2b^2
\left\lbrace(g^e_V + g^e_A)^2(p_1\cdot k_1)(p_2\cdot k_2)+ (g^e_V - g^e_A)^2 (p_1\cdot k_2) (p_2\cdot k_1)\right.\nonumber\\
&+&\left.\left[(g^e_V)^2-(g^e_A)^2\right] m_e^2(k_1\cdot k_2)\right\rbrace.
\end{eqnarray}

In the decoupling limit, when the mixing angle $\theta = 0$ and $a=b=1$, Eq. (12) is thus reduced to the expression
to the amplitude given in Refs. \cite{Yakovlev,Esposito1,Esposito2,Armando,Misiaszek,Dicus,Ellis}.

The stellar energy loss in the pair-annihilation process $e^+ +e^- \rightarrow \nu +\bar{\nu}$
is obtained by using Eq. (12). The formula of the stellar energy loss is given by
\cite{Yakovlev,Esposito1,Esposito2,Misiaszek,Lenard}

\begin{equation}
Q_{\nu\bar\nu}=\frac{4}{(2\pi)^8}\int\frac{d^3\mathbf{p}_1}{2E_1}\frac{d^3\mathbf{p}_2}{2E_2}
\frac{d^3\mathbf{k}_1}{2\epsilon_1}\frac{d^3\mathbf{k}_2}{2\epsilon_2}(E_1+E_2)F_1F_2
\delta^{(4)}(p_1+p_2-k_1-k_2)|{\cal M}|^2,
\end{equation}

\noindent where the quantities $F_{1,2}=[1+\exp(E_{e^-}\pm
\mu_{e^-} )/T]^{-1}$ are the Fermi-Dirac distribution functions
for $e^{\pm}$, $\mu_e$ is the chemical potential for the electrons
and $T$ is the temperature (we take $K_B=1$ for the Boltzmann
constant).

From the transition amplitude Eq. (12) and the formula of the stellar energy loss Eq. (13) we obtain

\begin{equation}
Q^{[1]}_{\nu\bar\nu}=8G_F^2a^2b^2\left(g^e_V + g^e_A\right)^2I_1,
\end{equation}

\noindent where $I_1$ is explicitly given by

\begin{equation}
I_1=\frac{4}{(2\pi)^8}\int\frac{d^3\mathbf{p}_1}{2E_1}\frac{d^3\mathbf{p}_2}{2E_2}
\frac{d^3\mathbf{k}_1}{2\epsilon_1}\frac{d^3\mathbf{k}_2}{2\epsilon_2}(E_1+E_2)F_1F_2
\delta^{(4)}(p_1+p_2-k_1-k_2)(p_1\cdot k_1)(p_2\cdot k_2).
\end{equation}

The integration can be performed by using the Lenard formula, namely \cite{Lenard}

\begin{eqnarray}
\int\frac{d^3\mathbf{k}_1}{2\epsilon_1}\frac{d^3\mathbf{k}_2}{2\epsilon_2}k_1^\alpha k_2^\beta\delta^{(4)}(p_1+p_2-k_1-k_2)
&=&\frac{\pi}{24}\left[g^{\alpha\beta}(p_1+p_2)^2+2(p_1^\alpha+p_2^\alpha)(p_1^\beta+p_2^\beta)\right]\nonumber\\
&&\cdot \Theta\left[(p_1+p_2)^2\right],
\end{eqnarray}

\noindent thus Eq. (15) takes the form

\begin{equation}
I_1=\frac{1}{24(2\pi)^7}\int\frac{d^3\mathbf{p}_1}{E_1}\frac{d^3\mathbf{p}_2}{E_2}(E_1+E_2)F_1F_2\left[
3m_e^2(p_1\cdot p_2)+2(p_1\cdot p_2)^2+m_e^4 \right ].
\end{equation}

Similarly for the second and third term of Eq. (12), we obtain

\begin{eqnarray}
Q^{[2]}_{\nu\bar\nu}&=&8G_F^2a^2b^2\left( g^e_V - g^e_A \right)^2I_2,\\
Q^{[3]}_{\nu\bar\nu}&=&8G_F^2a^2b^2\left[ (g^e_V)^2 - (g^e_A)^2\right]m^2_e I_3,
\end{eqnarray}

\noindent where

\begin{eqnarray}
I_2&=&I_1=\frac{1}{24(2\pi)^7}\int\frac{d^3\mathbf{p}_1}{E_1}\frac{d^3\mathbf{p}_2}{E_2}(E_1+E_2)F_1F_2
\left[3m_e^2(p_1\cdot p_2)+2(p_1\cdot p_2)^2+m_e^4 \right ],\\
I_3&=&\frac{1}{4(2\pi)^7}\int\frac{d^3\mathbf{p}_1}{E_1}\frac{d^3\mathbf{p}_2}{E_2}
(E_1+E_2)F_1F_2 \left[ (p_1\cdot p_2) + m_e^2 \right].
\end{eqnarray}

The calculation of the stellar energy loss rate can be more easily performed by expressing the latest integrals
in terms of the Fermi integral, which is defined  as \cite{Misiaszek}

\begin{equation}
G_s^{\pm}(\alpha, \beta, x)=\frac{1}{\alpha^{3+2s}}\int_\alpha^\infty
x^{2s+1}\frac{\sqrt{x^2-\alpha^2}}{1+e^{x\pm\beta}}dx,
\end{equation}

\noindent where $\alpha=\frac{m_e}{KT}$, $\beta=\frac{\mu_e}{KT}$ and $x=\frac{E}{KT}$.

With these definitions, Eq. (22) becomes

\begin{equation}
G_s^{\pm}=\frac{1}{m_e^{3+2s}}\int_{m_e/KT}^\infty
E^{2s+1}\frac{\sqrt{E^2- m_e^2}}{1+e^{(E\pm\mu_e)/KT}}dE,
\end{equation}

\noindent therefore

\begin{equation}
\int_{m_e/KT}^\infty E^{n}\frac{\sqrt{E^2-
m_e^2}}{1+e^{(E\pm\mu_e)/KT}}dE=m_e^{n+2}G_{\frac{n-1}{2}}^{\pm},
\end{equation}

\begin{equation}
\int_{m_e/KT}^\infty E^{n+1}\frac{\sqrt{E^2-
m_e^2}}{1+e^{(E\pm\mu_e)/KT}}dE=m_e^{n+3}G_{\frac{n}{2}}^{\pm},
\end{equation}

\begin{equation}
\int_{m_e/KT}^\infty E^{n+2}\frac{\sqrt{E^2-
m_e^2}}{1+e^{(E\pm\mu_e)/KT}}dE=m_e^{n+4}G_{\frac{n+1}{2}}^{\pm}.
\end{equation}

From (24)-(26), Eqs. (17), (20) and (21) are expressed as

\begin{eqnarray}
I_1^{nm}&=& I_2^{nm}= \frac{m_e^{n+m+8}}{6(2\pi)^5}\left[3G_{\frac{n}{2}}^{-}G_{\frac{m}{2}}^{+}+2G_{\frac{n+1}{2}}^{-}G_{\frac{m+1}{2}}^{+}
+G_{\frac{n-1}{2}}^{-}G_{\frac{m-1}{2}}^{+}\right. \nonumber\\
&&\left.
+\frac{4}{9}\left(G_{\frac{n+1}{2}}^{-}-G_{\frac{n-1}{2}}^{-}\right)\left(G_{\frac{m+1}{2}}^{+}-G_{\frac{m-1}{2}}^{+}\right)\right],\\
I_3^{nm}&=&\frac{m_e^{n+m+6}}{(2\pi)^5}\left[G_{\frac{n-1}{2}}^{-}G_{\frac{m-1}{2}}^{+}+G_{\frac{n}{2}}^{-}G_{\frac{m}{2}}^{+}\right].
\end{eqnarray}

Therefore, Eqs. (14), (18) and (19) are explicitly

\begin{eqnarray}
Q^{[1]}_{\nu\bar\nu}&=&8G_F^2a^2b^2\left[g^e_V + g^e_A\right]^2\left[ I_1^{10}+I_1^{01}\right],\\
Q^{[2]}_{\nu\bar\nu}&=&8G_F^2a^2b^2\left[g^e_V - g^e_A\right]^2\left[ I_2^{10}+I_2^{01}\right],\\
Q^{[3]}_{\nu\bar\nu}&=&8G_F^2\left[\left(g^e_V\right)^2- \left( g^e_A\right)^2\right]m_e^2\left[I_3^{10}+I_3^{01}\right].
\end{eqnarray}

Finally, the expression for the stellar energy loss of neutrino pair production is given by

\begin{equation}
Q^{331}_{\nu\bar\nu}\left(\theta,\beta\right)=Q_{\nu\bar\nu}^{[1]}\left(\theta,\beta\right)+Q_{\nu\bar\nu}^{[2]}\left(\theta,\beta\right)
+Q_{\nu\bar\nu}^{[3]}\left(\theta,\beta\right),
\end{equation}

\noindent this is an exact result for all values of the $\alpha$ and $\beta$, i.e., whether or not the electrons
are degenerate or relativistic.

We emphasise that the dependence of the mixing angle $\theta$ between $Z-Z'$ of the SM and the 331M is
contained in the constants $a$ and $b$, while the dependence of the $\beta$ degeneration parameter is
contained in the Fermi integrals $G_s^{\pm}(\alpha, \beta, x)$, respectively.

It is noteworthy that the Fermi integrals $G_s^{\pm}(\alpha, \beta, x)$ given in Eq. (22) can not be done
analytically for all $\alpha$ and $\beta$, i.e., we cannot find an analytic expression for
$Q^{331}_{\nu\bar\nu}\left(\theta,\beta\right)$ which holds for all values of temperature $T$ and chemical
potential $\mu_e$. However, with the purpose of comparison our new contribution with the standard result, we will
evaluate Eq. (32) in various limits regions of $\alpha=\frac{m_e}{KT}$ and $\beta=\frac{\mu_e}{KT}$. In addition,
to see the effects of $\theta$, the free parameter of the 331M, as well as the deviation of the stellar energy
loss rate in our model from the standard one, we define the relative correction

\begin{equation}
\frac{\delta Q}{Q^{SM}_{\nu\bar \nu}}= \frac{Q^{331}_{\nu\bar\nu}\left(\theta,\beta\right)- Q^{SM}_{\nu\bar\nu}\left(\beta\right)}{Q^{SM}_{\nu\bar\nu}\left(\beta\right)},
\end{equation}

\noindent as a function of $\theta$ and $\beta$. Having done this we obtain the relative correction as follows.\\

\noindent {\bf Region I}: In this nonrelativistic and nondegenerate case ($1\ll \alpha$, $\beta \ll \alpha$)
characterized by temperatures between $3\times 10^8 \leq T \leq3\times 10^9\hspace{1mm}^oK$ and density  $\rho \leq 10^5\hspace{0.8mm}g/cm^3$,
with higher densities requiring higher temperatures.

For the Fermi integrals given in Eq. (22) we make the variable change $z=x-\alpha$.
Therefore,

\begin{eqnarray}
x&=&z+\alpha,\nonumber\\
x^2-\alpha^2&=&z^2+2z\alpha,\\
dx&=&dz,\nonumber
\end{eqnarray}

\noindent and for this new variable, the integration limits change from 0 to $\infty$. Thus

\begin{equation}
G^{\pm}_n=\sqrt{2}\alpha^{-3/2}\int ^\infty_0\frac{(\alpha^{-1} z+1)^{2n+1}z^{1/2}(1+\frac{\alpha^{-1} z}{2})^{1/2}}{1+e^{z+\alpha\pm \beta}}dz,
\end{equation}

\noindent and applying the approximation $\beta \ll \alpha$, we get

\begin{equation}
G^{\pm}_n=\sqrt{2}\frac{\alpha^{-3/2}}{e^{\alpha\pm \beta}}\int ^\infty_0(\alpha^{-1} z+1)^{2n+1}z^{1/2}
(1+\frac{\alpha^{-1} z}{2})^{1/2}e^{-z}dz.
\end{equation}

Now, applying the condition $1 \ll \alpha$, we see that for large $z$, we always can find a $\alpha^{-1}$
such that $0< 1 \ll \alpha$, and $z \ll \alpha$. Therefore,

\begin{equation}
G^{\pm}_n=\sqrt{2}\alpha^{-3/2} e^{-\alpha \mp \beta}\int ^\infty_0[1+(2n+1)\alpha^{-1} z]z^{1/2}
(1+\frac{\alpha^{-1} z}{4})e^{-z}dz,
\end{equation}

\noindent from which, the quadratic terms can be neglected to give us,

\begin{eqnarray}
G^{\pm}_n&=&\sqrt{2}\alpha^{-3/2} e^{-\alpha} e^{\mp \beta}\int ^\infty_0[1+(2n+\frac{5}{4})\alpha^{-1} z]z^{1/2}e^{-z}dz,\nonumber\\
&=&\sqrt{2}\alpha^{-3/2} e^{-\alpha} e^{\mp \beta} \left[\int^\infty_0 z^{1/2}e^{-z}dz
+ (2n+\frac{5}{4})\alpha^{-1} \int^\infty_0 z^{3/2}e^{-z}dz \right],\\
&=&\sqrt{2}\alpha^{-3/2} e^{-\alpha} e^{\mp \beta} \left[\Gamma(3/2)+(2n+\frac{5}{4})\alpha^{-1} \Gamma(5/2)\right],\nonumber
\end{eqnarray}

\noindent finally, we obtain

\begin{equation}
G^{\pm}_n=\sqrt{\frac{\pi}{2}}\alpha^{-3/2} e^{-\alpha} e^{\mp \beta} \left[1+\frac{3}{2}(2n+\frac{5}{4})\alpha^{-1} \right].
\end{equation}

So, a first approximation is given by

\begin{equation}
G^{\pm}_0=\sqrt{\frac{\pi}{2}}\alpha^{-3/2} e^{-\alpha} e^{\mp \beta} \left(1+\frac{15}{8}\alpha^{-1} \right).
\end{equation}

Now, taking into account the sign, the condition $1 \ll \alpha$, and the first order of result (40), we get

\begin{equation}
G^{\pm}_n\approx G^{\pm}_0 =\sqrt{\frac{\pi}{2}}\alpha^{-3/2} e^{-\alpha} e^{\mp \beta}.
\end{equation}

With these approximations and after of a direct calculation we get the relative correction for the region I

\begin{equation}
\frac{\delta Q_I}{Q^{SM}_I}=
\frac{a^2b^2\left[(g^e_V)^2 +(g^e_A)^2\right]- \left[(g^e_V)^2 +(g^e_A)^2\right]}{\left[(g^e_V)^2 +(g^e_A)^2\right]}.
\end{equation}

\noindent {\bf Region II}: For the nonrelativistic and mildly degenerate case ($1 \ll \alpha$, $\alpha \ll \beta\ll 2\alpha$), the
temperature $T< 10^8\hspace{0.8mm}^oK$.

\noindent In this case, it holds that $G^-_0\gg G^+_0$ and $G^-_n\approx G^-_0$, so that the result for $G^\pm_n$ given by Eq. (41)
remains valid, and we obtain

\begin{equation}
\frac{\delta Q_{II}}{Q^{SM}_{II}}=
\frac{a^2b^2\left[(g^e_V)^2 +(g^e_A)^2\right]- \left[(g^e_V)^2 +(g^e_A)^2\right]}{\left[(g^e_V)^2 +(g^e_A)^2\right]}.
\end{equation}

\noindent {\bf Region III}: Relativistic and degenerate case ($1 \ll \alpha$, $\beta \gg \alpha$), valid for temperatures
$T > 6\times 10^7\hspace{0.8mm}^oK$ and densities $\rho > 10^7\hspace{0.8mm}g/cm^3$.

From the condition $1 \ll \alpha$, the following is obtained

\begin{equation}
G^+_n\approx G^+_0 =\sqrt{\frac{\pi}{2}}\alpha^{-3/2} e^{-\alpha} e^{- \beta},
\end{equation}

\noindent and of the condition $\beta \gg \alpha$ it holds that $G^-_n \gg G^+_n$.

In addition, of the Fermi integral $G^-_n$ and using the condition $\beta \gg \alpha$

\begin{eqnarray}
G_n^-(\alpha, \beta, x)&=&\frac{1}{\alpha^{3+2n}}\int_\alpha^\infty x^{2n+1}\frac{\sqrt{x^2-\alpha^2}}{1+e^{x\pm\beta}}dx,\nonumber\\
                        &\approx&\frac{1}{\alpha^{3+2n}}\int_\alpha^\infty x^{2n+1}\sqrt{x^2-\alpha^2}dx,
\end{eqnarray}

\noindent integrating by parts repeatedly and using the condition $\beta \gg \alpha$, one can show that

\begin{equation}
G^-_n\approx \frac{3}{2n+3}(\alpha^{-1}\beta)^{2n}G^-_0.
\end{equation}

In this case the relative correction is given by

\begin{equation}
\frac{\delta Q_{III}}{Q^{SM}_{III}}=
\frac{a^2b^2\left[(g^e_V)^2 +(g^e_A)^2\right]- \left[(g^e_V)^2 +(g^e_A)^2\right]}{\left[(g^e_V)^2 +(g^e_A)^2\right]}.
\end{equation}

\noindent {\bf Region IV}: The relativistic and nondegenerate case ($\alpha \ll 1$, $\beta \ll 1$), is for densities $\rho > 10^7\hspace{0.8mm}g/cm^3$.

In this case, the Fermi integrals can be approximated as

\begin{equation}
G_n^- \approx \frac{1}{\alpha^{3+2n}}\int_\alpha^\infty \frac{x^{2n+2}}{1+e^{x\pm\beta}}dx,
\end{equation}

\noindent using

\[ \frac{x^{2n+2}}{1+e^{x\pm\beta}}= x^{2n+2}\sum^\infty_{k=0} (-1)^k e^{k(x\pm\beta)},\]

\noindent and applying the condition $\beta \ll 1$, the Fermi integrals are

\begin{equation}
\int_\alpha^\infty \frac{x^{2n+2}}{1+e^{x\pm\beta}}dx=\sum^\infty_{k=0} \int_\alpha^\infty e^{kx}x^{2n+2}dx.
\end{equation}

\noindent After integrating it follows that

\begin{equation}
G^{\pm}_n\approx \frac{1}{\alpha^{3+2n}}\Gamma{(2n+3)}\eta(2n+3),
\end{equation}

\noindent where $\Gamma(2n+3)$ is the gamma function, while $\eta(2n+3)$ is defined by

\begin{equation}
\eta(2n+3)=\sum^\infty_0\frac{(-1)^{k+1}}{k^{2n+3}}.
\end{equation}

Therefore, in this case we obtain

\begin{equation}
\frac{\delta Q_{IV}}{Q^{SM}_{IV}}=
\frac{a^2b^2\left[(g^e_V)^2 +(g^e_A)^2\right]- \left[(g^e_V)^2 +(g^e_A)^2\right]}{\left[(g^e_V)^2 +(g^e_A)^2\right]}.
\end{equation}

\noindent {\bf Region V}: For the relativistic and degenerate case ($\alpha \ll 1$, $\beta \gg 1$), the restriction is for temperatures
and densities of $T= 10^{10}\hspace{0.8mm}^oK$ and $\rho > 10^8\hspace{0.8mm}g/cm^3$.

From the condition $\alpha \ll 1$, the Fermi integral $G^+_n$ can be approximated as

\begin{equation}
G_n^+ \approx \frac{1}{\alpha^{3+2n}}\int_\alpha^\infty \frac{x^{2n+2}}{1+e^{x +\beta}}dx,
\end{equation}

\noindent and of the condition $\beta \gg 1$ is obtained

\begin{eqnarray}
G_n^+ &\approx& \frac{1}{\alpha^{3+2n}}\int_\alpha^\infty \frac{x^{2n+2}}{e^{x +\beta}}dx,\nonumber\\
      &=& \frac{1}{\alpha^{3+2n}} e^{-\beta}\Gamma(2n+3),\\
      &=& \frac{1}{\alpha^{3+2n}} e^{-\beta}(2n+2)!.\nonumber
\end{eqnarray}

In addition, we have

\begin{equation}
G_n^- \approx \frac{(\alpha^{-1}\beta)^{2n+3}}{2n+3},
\end{equation}

\noindent and

\begin{equation}
G_0^- \approx \frac{(\alpha^{-1}\beta)^3}{3},
\end{equation}

\noindent therefore,

\begin{equation}
G_n^- \approx \frac{(3\alpha^{-1}\beta)^{2n}}{2n+3}G^-_0.
\end{equation}

In this case the relative correction is given by

\begin{equation}
\frac{\delta Q_{V}}{Q^{SM}_{V}}=
\frac{a^2b^2\left[(g^e_V)^2 +(g^e_A)^2\right]- \left[(g^e_V)^2 +(g^e_A)^2\right]}{\left[(g^e_V)^2 +(g^e_A)^2\right]}.
\end{equation}

In general, the relative correction for the stellar energy loss rate for the different regions I-V is given by

\begin{equation}
\frac{\delta Q_{I-V}}{Q^{SM}_{I-V}}=
\frac{a^2b^2\left[(g^e_V)^2 +(g^e_A)^2\right]- \left[(g^e_V)^2 +(g^e_A)^2\right]}{\left[(g^e_V)^2 +(g^e_A)^2\right]}.
\end{equation}

\subsection{Stellar energy loss rates through $e^+ +e^- \rightarrow \nu +\bar{\nu}$ in a left-right symmetric model}

Another potentially interesting model, is the LRSM \cite{Pati,Mohapatra0,Mohapatra1,Senjanovic,Senjanovic1,Polak,A.Gutierrez6}.
In the context from this model, the amplitude of transition for the process (5) is given by

\begin{equation}
{\cal M}=\frac{g^2}{2M_Z^2}\left[\bar{u}\left(k_2,\lambda_2\right)\gamma^\mu\frac{1}{2}\left(a' g_V^\nu-b' g_A^\nu
\gamma_5\right)v\left(k_1,\lambda_1\right)\right]\left[\bar{v}\left(p_1\right)\gamma_\mu\frac{1}{2}\left(a'
g_V^e-b' g_A^e \gamma_5\right)u\left(p_2\right)\right],
\end{equation}

\noindent where the constant $a'$ and $b'$ depend only on the parameters of the LRSM \cite{Gutierrez0}

\begin{equation} a'=\cos{\phi}-\frac{\sin{\phi}}{\sqrt{\cos2\theta_W}} \hspace{5mm} \mbox{and}
\hspace{5mm} b'=\cos{\phi}+\sqrt{\cos2\theta_W}\sin{\phi},
\end{equation}

\noindent and $\phi$ is the mixing angle $Z-Z'$ of the LRSM.

The explicit form for the squared transition amplitude is

\begin{eqnarray}
\sum_s|{\cal M}|^2=&4&G_F^2
(a'^2+b'^2)\left\lbrace\left[ a'^2(g^e_V)^2+b'^2(g^e_A)+4\frac{a'^2b'^2}{(a'^2+b'^2)}g^e_Vg^e_A\right] (p_1\cdot k_1) (p_2\cdot k_2)\right.\nonumber\\
&+&\left[ a'^2(g^e_V)^2+b'^2(g^e_A)-4\frac{a'^2b'^2}{(a'^2+b'^2)}g^e_Vg^e_A\right] (p_1\cdot k_2)(p_2\cdot k_1)\nonumber\\
&+&\left.\left[a'^2(g^e_V)^2-b'^2(g^e_A)\right]m_e^2(k_1\cdot k_2)\right\rbrace.
\end{eqnarray}

In the decoupling limit when the mixing angle $\phi = 0$ and $a'=b'=1$, Eq. (62) is thus reduced to the expression to the amplitude given in
literature \cite{Yakovlev,Esposito1,Esposito2,Armando,Misiaszek,Dicus,Ellis}.

To derive the expression for the stellar energy loss rates, we follow the methodology as in subsection A and make the respective changes
to get

\begin{equation}
Q^{LRSM}_{\nu\bar\nu}\left(\phi,\beta\right)=Q_{\nu\bar\nu}^{[1]}\left(\phi,\beta\right)+Q_{\nu\bar\nu}^{[2]}\left(\phi,\beta\right)
+Q_{\nu\bar\nu}^{[3]}\left(\phi,\beta\right),
\end{equation}

\noindent with

\begin{eqnarray}
Q^{[1]}_{\nu\bar\nu}&=&4G_F^2(a'^2+b'^2)\left[a'^2\left(g^e_V\right)^2+b'^2 \left( g^e_A\right)^2+\frac{4a'^2b'^2}{a'^2+b'^2}g^e_V g^e_A\right]
\left[ I_1^{10}+I_1^{01}\right],\\
Q^{[2]}_{\nu\bar\nu}&=&4G_F^2(a'^2+b'^2)\left[a'^2\left(g^e_V\right)^2+b'^2 \left( g^e_A\right)^2-\frac{4a'^2b'^2}{a'^2+b'^2}g^e_V g^e_A\right]
\left[ I_2^{10}+I_2^{01}\right],\\
Q^{[3]}_{\nu\bar\nu}&=&4G_F^2(a'^2+b'^2)\left[a'^2\left(g^e_V\right)^2-b'^2 \left( g^e_A\right)^2\right]m_e^2\left[
I_3^{10}+I_3^{01}\right],
\end{eqnarray}

\noindent where the dependence of the $\beta$ degeneration parameter is contained in the Fermi integrals $G_s^{\pm}(\alpha, \beta, x)$.

As mentioned in the Subsecti\'on A, Fermi integrals $G_s^{\pm}(\alpha, \beta, x)$ which are given by Eq. (22) can not be solved analytically,
but only for certain cases limits of the temperature $T$ and chemical potential. For this reason, we consider the following approximations.\\

\noindent {\bf Region I}: The nonrelativistic and nondegenerate case ($1 \ll \alpha $, $\beta \ll \alpha$), is characterized by temperatures
between $3\times 10^8 \leq T \leq 3\times 10^9\hspace{1mm}^oK$ and density  $\rho \leq 10^5\hspace{0.8mm}g/cm^3$. Higher densities requiring higher temperatures, thus we get

\begin{equation}
\frac{\delta Q_I}{Q^{SM}_I}=
\frac{a'^2(a'^2 + b'^2)(g^e_V)^2 - 2(g^e_V)^2}{2(g^e_V)^2}.
\end{equation}

\noindent {\bf Region II}: In the nonrelativistic and mildly degenerate case ($1\ll \alpha$, $\alpha \ll \beta\ll 2\alpha$) and
$T< 10^8\hspace{0.8mm}^oK$ we obtain

\begin{equation}
\frac{\delta Q_{II}}{Q^{SM}_{II}}=
\frac{a'^2(a'^2 + b'^2)(g^e_V)^2 - 2(g^e_V)^2}{2(g^e_V)^2}.
\end{equation}

\noindent {\bf Region III}: The relativistic and degenerate case ($1\ll \alpha$, $\beta \gg \alpha$), with temperatures and densities
of $T > 6\times 10^7\hspace{0.8mm}^oK$ and $\rho > 10^7\hspace{0.8mm}g/cm^3$. The relative correction is given by

\begin{equation}
\frac{\delta Q_{III}}{Q^{SM}_{III}}=
\frac{(a'^2+b'^2)\left[a'^2(g^e_V)^2 +b'^2(g^e_A)^2\right]- 2\left[(g^e_V)^2 +(g^e_A)^2\right]}{2\left[(g^e_V)^2 +(g^e_A)^2\right]}.
\end{equation}

\noindent {\bf Region IV}: For the relativistic and nondegenerate case ($ \alpha \ll 1$, $\beta \ll 1$) with densities $\rho > 10^7\hspace{0.8mm}g/cm^3$.
In this case the relative correction is

\begin{equation}
\frac{\delta Q_{IV}}{Q^{SM}_{IV}}=
\frac{(a'^2+b'^2)\left[a'^2(g^e_V)^2 +b'^2(g^e_A)^2\right]- 2\left[(g^e_V)^2 +(g^e_A)^2\right]}{2\left[(g^e_V)^2 +(g^e_A)^2\right]}.
\end{equation}

\noindent {\bf Region V}: Relativistic and degenerate case ($\alpha \ll 1$, $\beta \gg 1$), the restricted
is for temperatures $T= 10^{10}\hspace{0.8mm}^oK$ and densities $\rho > 10^8\hspace{0.8mm}g/cm^3$ obtaining

\begin{equation}
\frac{\delta Q_{V}}{Q^{SM}_{V}}=
\frac{(a'^2+b'^2)\left[a'^2(g^e_V)^2 +b'^2(g^e_A)^2\right]- 2\left[(g^e_V)^2 +(g^e_A)^2\right]}{2\left[(g^e_V)^2 +(g^e_A)^2\right]}.
\end{equation}

Finally, we summarize the relative correction as follows:

\begin{equation}
\frac{\delta Q_{I-II}}{Q^{SM}_{I-II}}=
\frac{a'^2(a'^2 + b'^2)(g^e_V)^2 - 2(g^e_V)^2}{2(g^e_V)^2},
\end{equation}

\noindent and

\begin{equation}
\frac{\delta Q_{III-V}}{Q^{SM}_{III-V}}=
\frac{(a'^2+b'^2)\left[a'^2(g^e_V)^2 +b'^2(g^e_A)^2\right]- 2\left[(g^e_V)^2 +(g^e_A)^2\right]}{2\left[(g^e_V)^2 +(g^e_A)^2\right]}.
\end{equation}

\subsection{Stellar energy loss rates through $e^+ +e^- \rightarrow \nu +\bar{\nu}$ in a simplest little Higgs model}

In this subsection we calculate the stellar energy loss rate through the reaction $e^{+}e^{-}\rightarrow \nu \bar\nu$
using the neutral current lagrangian given in Eq. (20) of Ref. \cite{Dias} for the SLHM. A interesting characteristic
from this model is that is independent of the mass of the additional $Z_H$ heavy gauge boson and so we have the characteristic
energy scale of the model $f$ as the only additional parameter. The respective transition amplitude is given by

\begin{equation}
{\cal M}=\frac{g^2}{M^2_Z\cos^2\theta_W}\left[\bar{u}\left(k_2,\lambda_2\right)\gamma^\mu\frac{1}{2}\left( g_V^\nu- g_A^\nu
\gamma_5\right)v\left(k_1,\lambda_1\right)\right]\left[\bar{v}\left(p_1\right)\gamma_\mu\frac{1}{2}\left( g_V^e- g_A^e \gamma_5\right)u\left(p_2\right)\right],
\end{equation}

\noindent where explicitly the coupling constants $g^e_V (g^\nu_V)$ and $g^e_A (g^\nu_A)$ which contain the
characteristic energy scale $f$ of the SLHM are

\begin{eqnarray}
g^e_V&=&-\left(\frac{1}{2}-2\sin^2\theta_W \right) \left( 1-\left(\frac{ 1-4\cos^2\theta_W}{8\cos^4\theta_W}\right)\frac{v^2}{f^2} \right),\nonumber\\
g^e_A&=&-\frac{1}{2} +\left(\frac{1-4\cos^2\theta_W}{16\cos^4\theta_W}\right)\frac{v^2}{f^2},\nonumber\\
g^\nu_V&=&\frac{1}{2} -\left(\frac{1-4\cos^2\theta_W}{16\cos^4\theta_W}\right)\frac{v^2}{f^2},\\
g^\nu_A&=&\frac{1}{2} +\left(\frac{1-4\cos^2\theta_W}{16\cos^4\theta_W}\right)\frac{v^2}{f^2}.\nonumber
\end{eqnarray}

After making the corresponding algebra, the explicit expression for the square of the transition amplitude is

\begin{eqnarray}
\sum_s|{\cal M}|^2=&64&G_F^2
\left[(g^\nu_V)^2+(g^\nu_A)^2\right]\left\lbrace\left[(g^e_V)^2 + (g^e_A)^2+\frac{4g^\nu_V g^\nu_A g^e_V g^e_A}{\left( (g^\nu_V)^2 + (g^\nu_A)^2\right)}\right] (p_1\cdot k_1) (p_2\cdot k_2)\right.\nonumber\\
&+&\left[(g^e_V)^2+(g^e_A)^2-\frac{4g^\nu_V g^\nu_A g^e_V g^e_A}{\left( (g^\nu_V)^2 + (g^\nu_A)^2\right)}\right](p_1\cdot k_2)(p_2\cdot k_1)\nonumber\\
&+&\left.\left[(g^e_V)^2-(g^e_A)^2\right]m_e^2(k_1\cdot k_2)\right\rbrace.
\end{eqnarray}

The stellar energy loss rates through $e^+ +e^- \rightarrow \nu +\bar{\nu}$ in a SLHM is given by

\begin{equation}
Q^{SLHM}_{\nu\bar\nu}\left(f,\beta\right)=Q_{\nu\bar\nu}^{[1]}\left(f,\beta\right)+Q_{\nu\bar\nu}^{[2]}\left(f,\beta\right)
+Q_{\nu\bar\nu}^{[3]}\left(f,\beta\right),
\end{equation}

\noindent where

\begin{eqnarray}
Q^{[1]}_{\nu\bar\nu}&=&16G_F^2
\left[(g^\nu_V)^2+(g^\nu_A)^2\right]\left[(g^e_V)^2 + (g^e_A)^2+\frac{4g^\nu_V g^\nu_A g^e_V g^e_A}{\left( (g^\nu_V)^2 + (g^\nu_A)^2\right)}\right]
\left[ I_1^{10}+I_1^{01}\right],\\
Q^{[2]}_{\nu\bar\nu}&=&16G_F^2
\left[(g^\nu_V)^2+(g^\nu_A)^2\right]\left[(g^e_V)^2 + (g^e_A)^2-\frac{4g^\nu_V g^\nu_A g^e_V g^e_A}{\left( (g^\nu_V)^2 + (g^\nu_A)^2\right)}\right]
\left[ I_2^{10}+I_2^{01}\right]\\
Q^{[3]}_{\nu\bar\nu}&=&16G_F^2
\left[(g^\nu_V)^2+(g^\nu_A)^2\right]\left[(g^e_V)^2-(g^e_A)^2\right]m_e^2\left[I_3^{10}+I_3^{01}\right].
\end{eqnarray}

To study the effects of the scale of energy $f$, which is the free parameter of the SLHM with respect to the standard result
we consider different limiting cases for $\alpha=\frac{m_e}{KT}$ and $\beta=\frac{\mu_e}{KT}$. For this, we consider the
relative correction which is defined as in Eq. (33). The following cases are considered:\\

\noindent {\bf Region I}: In the nonrelativistic and nondegenerate case ($1\ll \alpha$, $\beta \ll \alpha$) and
characterized by temperatures between $3\times 10^8\leq T \leq 3\times 10^9\hspace{1mm}^oK$ and density
$\rho \leq 10^5\hspace{0.8mm}g/cm^3$, we get

\begin{equation}
\frac{\delta Q_I}{Q^{SM}_I}=
\frac{2(g^e_V)^2\left[(g^\nu_V)^2 +(g^\nu_A)^2\right]-(g^{eSM}_V)^2}{(g^{eSM}_V)^2},
\end{equation}

\noindent where the parameter of scale $f$ of the SLHM is contained in the constants $g^e_V (g^\nu_V)$ and $g^e_A (g^\nu_A)$
defined in Eq. (75).\\

\noindent {\bf Region II}: The nonrelativistic and mildly degenerate case ($1 \ll \alpha$, $\alpha \ll \beta\ll 2\alpha$) is for
$T< 10^8\hspace{0.8mm}^oK$. In this case, the relative correction is

\begin{equation}
\frac{\delta Q_{II}}{Q^{SM}_{II}}=
\frac{2(g^e_V)^2\left[(g^\nu_V)^2 +(g^\nu_A)^2\right]-(g^{eSM}_V)^2}{(g^{eSM}_V)^2}.
\end{equation}

\noindent {\bf Region III}: Relativistic and degenerate case ($1 \ll \alpha$, $\beta \gg \alpha$), this region is for temperatures
$T > 6\times 10^7\hspace{0.8mm}^oK$ and densities $\rho > 10^7\hspace{0.8mm}g/cm^3$. The relative correction is given by

\begin{equation}
\frac{\delta Q_{III}}{Q^{SM}_{III}}=
\frac{2\left[(g^\nu_V)^2 +(g^\nu_A)^2\right]\left[(g^e_V)^2 +(g^e_A)^2\right]- \left[(g^{eSM}_V)^2 +(g^{eSM}_A)^2\right]}{\left[(g^{eSM}_V)^2 +(g^{eSM}_A)^2\right]}.
\end{equation}

\noindent {\bf Region IV}: The relativistic and nondegenerate case ($\alpha \ll 1$, $\beta \ll 1$), is for
densities $\rho > 10^7\hspace{0.8mm}g/cm^3$ and the correspond expression for the relative correction is

\begin{equation}
\frac{\delta Q_{IV}}{Q^{SM}_{IV}}=
\frac{2\left[(g^\nu_V)^2 +(g^\nu_A)^2\right]\left[(g^e_V)^2 +(g^e_A)^2\right]- \left[(g^{eSM}_V)^2 +(g^{eSM}_A)^2\right]}{\left[(g^{eSM}_V)^2 +(g^{eSM}_A)^2\right]}.
\end{equation}

\noindent {\bf Region V}: The relativistic and degenerate case ($\alpha \ll 1$, $\beta \gg 1$), is restricted
at temperatures $T= 10^{10}\hspace{0.8mm}^oK$ and densities $\rho > 10^8\hspace{0.8mm}g/cm^3$ obtaining

\begin{equation}
\frac{\delta Q_{V}}{Q^{SM}_{V}}=
\frac{2\left[(g^\nu_V)^2 +(g^\nu_A)^2\right]\left[(g^e_V)^2 +(g^e_A)^2\right]- \left[(g^{eSM}_V)^2 +(g^{eSM}_A)^2\right]}{\left[(g^{eSM}_V)^2 +(g^{eSM}_A)^2\right]}.
\end{equation}

In summary, the relative correction for the stellar energy loss rates for the regions I and II is given by

\begin{equation}
\frac{\delta Q_{I-II}}{Q^{SM}_{I-II}}=
\frac{2(g^e_V)^2\left[(g^\nu_V)^2 +(g^\nu_A)^2\right]-(g^{eSM}_V)^2}{(g^{eSM}_V)^2},
\end{equation}

\noindent while in the case of the regions III-V we obtain

\begin{equation}
\frac{\delta Q_{III-V}}{Q^{SM}_{III-V}}=
\frac{2\left[(g^\nu_V)^2 +(g^\nu_A)^2\right]\left[(g^e_V)^2 +(g^e_A)^2\right]- \left[(g^{eSM}_V)^2 +(g^{eSM}_A)^2\right]}{\left[(g^{eSM}_V)^2 +(g^{eSM}_A)^2\right]}.
\end{equation}

\section{Results and Conclusions}

An comprehensive calculation of the stellar energy loss rates through the neutrino pair production via the
process $e^+ e^- \to \nu \bar \nu$ in the context of a 331M, a LRSM and the SLHM as a function of the
degeneration parameter $\beta$, as well as, of the parameters of each model, the mixing angles $\theta$,
$\phi$ and the energy scale $f$, has been addressed.

\begin{table}[!ht]
\caption{Physical constants \cite{Data2014}.}
\begin{center}
 \begin{tabular}{|c|c|}
\hline\hline
Quantity                  &                 Value                              \\
\hline
Electron mass             &       $m_e= 0.510998928\pm 0.000000011$ $MeV$       \\

Gauge boson mass          &       $M_Z= 91.1876\pm 0.0021$ $GeV$       \\

Fermi constant   &       $G_F=1.166 378 7(6)\times10^{-5}$ $GeV^{-2}$   \\

Weak mixing angle         &       $\sin^2\theta_W=0.23149\pm 0.00016$            \\

Expectation value of the vacuum   & $v=246\hspace{0.8mm}GeV$                      \\
\hline\hline
\end{tabular}
\end{center}
\end{table}

For the numerical calculation we have considered the input data \cite{Data2014} given in Table I, thereby obtaining the stellar
energy loss rates of the neutrinos $Q^{331}_{\nu\bar\nu}=Q^{331}_{\nu\bar\nu}\left(\theta,\beta\right)$,
$Q^{LRSM}_{\nu\bar\nu}=Q^{LRSM}_{\nu\bar\nu}\left(\phi,\beta\right)$ and $Q^{SLHM}_{\nu\bar\nu}=Q^{SLHM}_{\nu\bar\nu}\left(f,\beta\right)$.

For the mixing angle $Z-Z'$ of the 331M \cite{Cogollo} and LRSM \cite{A.Gutierrez2} we consider the following

\begin{eqnarray}
-3.979\times 10^{-3}\leq &\theta& \leq 1.309\times 10^{-4}, \hspace{5mm} \mbox{90\% C.L.},\\
-1.6\times 10^{-3}\leq &\phi& \leq 1.1\times 10^{-3}, \hspace{5mm} \mbox{90\% C.L.}
\end{eqnarray}

Other limits on the mixing angles $\theta$ and $\phi$ reported in the literature are given in Refs.
\cite{Long,Gutierrez5} and \cite{A.Gutierrez3,Polak,Gutierrez0,Adriani}. While for the characteristic energy scale $f$ of
the SLHM we consider

\begin{equation}
1.5\leq f \leq 10 \hspace{0.8mm}TeV,
\end{equation}

\noindent there are other limits on $f$ reported in Refs. \cite{Dias,Marandella,A.Gutierrez4}.

In Fig. 1 we show the stellar energy loss rates $Q^{331}_{\nu\bar \nu}(\theta, \beta)$ as a function
of the degeneracy parameter $\beta$ and different values of the mixing angle $\theta=-3.979\times 10^{-3},\hspace{0.8mm} 0,\hspace{0.8mm} 1.309\times 10^{-4}$,
which is defined by Eq. (32). We observe that the stellar energy loss rates remains almost constant for any value of the mixing angle
$\theta$ and decreases when $\beta$ increases, which is due to the reduction in the number of positrons available necessary to cause
the collision.

To visualize the effects of $\theta$, the free parameter of the 331M on the stellar energy loss rates
we plot the relative correction for the different regions which were already discussed in the text

\begin{equation}
\frac{\delta Q_{I-V}}{Q^{SM}_{I-V}}=
\frac{a^2b^2\left[(g^e_V)^2 +(g^e_A)^2\right]- \left[(g^e_V)^2 +(g^e_A)^2\right]}{\left[(g^e_V)^2 +(g^e_A)^2\right]},
\end{equation}

\noindent as a function of $\theta$ and $\sin^2\theta_W=0.23149-0.00016,\hspace{0.8mm} 0.23149,\hspace{0.8mm} 0.23149+0.00016$, in Fig. 2. We can
see that the relative correction  reaches its maximum value for the lower limit of $\theta$ and decreases as $\theta$
increases, remaining constant with respect to $\sin^2\theta_W$. The relative correction is of the order of $0.4\%$ relative
to the value of the standard model \cite{Dicus}.

In the case of the LRSM, we plot the stellar energy loss rates as a function of $\beta$ and of the mixing angle $Z-Z'$ of the model,
that is to say $\phi=-1.6\times 10^{-3},\hspace{0.8mm} 0,\hspace{0.8mm} 1.1\times 10^{-3}$, in Fig. 3. The $Q^{LRSM}_{\nu\bar \nu}(\phi, \beta)$ has a very
similar behaviour as in the case of the 331 model, this is due to the fact that the mixing angles from these models are very restricted
and both are of the same order of magnitude.

The deviation of the stellar energy loss rates in the LRSM from the SM one, for the regions I-II and III-V

\begin{equation}
\frac{\delta Q_{I-II}}{Q^{SM}_{I-II}}=
\frac{a'^2(a'^2 + b'^2)(g^e_V)^2 - 2(g^e_V)^2}{2(g^e_V)^2},
\end{equation}

\noindent and

\begin{equation}
\frac{\delta Q_{III-V}}{Q^{SM}_{III-V}}=
\frac{(a'^2+b'^2)\left[a'^2(g^e_V)^2 +b'^2(g^e_A)^2\right]- 2\left[(g^e_V)^2 +(g^e_A)^2\right]}{2\left[(g^e_V)^2 +(g^e_A)^2\right]},
\end{equation}

\noindent are depicted in Fig. 4 as a function of the parameter of mixing $\phi$ and different values of
the $\sin^2\theta_W=0.23149-0.00016,\hspace{0.8mm} 0.23149,\hspace{0.8mm} 0.23149+0.00016$ of the Weinberg angle.
Fig. 4 shows that the relative correction is sensitive to the mixing angle $\phi$, however it is independent of $\sin^2\theta_W$.
From this figure we observed that the relative correction $\frac{\delta Q_{I-II}}{Q^{SM}_{I-II}}$ is of the order of $0.5\%$ to the
lower bound of $\phi$ given in Eq. (90), whereas for $\frac{\delta Q_{III-V}}{Q^{SM}_{III-V}}$ is of the order of $0.2\%$ for the
lower and upper bounds of the mixing angle.

In Fig. 5 we show the stellar energy loss rate $Q^{SLHM}_{\nu\bar \nu}(f, \beta)$. It is noteworthy mentioning that the curves obtained
are very similar to those obtained in the 331 model and the LRSM.

Finally, to analyze the contribution of the energy scale $f$ of the SLHM on the stellar energy loss rates
$Q^{SLHM}_{\nu\bar\nu}\left(f, \beta\right)$ of the neutrinos, in Fig. 6 we show the relative change for the regions I and II

\begin{equation}
\frac{\delta Q_{I-II}}{Q^{SM}_{I-II}}=
\frac{2(g^e_V)^2\left[(g^\nu_V)^2 +(g^\nu_A)^2\right]-(g^{eSM}_V)^2}{(g^{eSM}_V)^2},
\end{equation}

\noindent as well as for the regions III-V

\begin{equation}
\frac{\delta Q_{III-V}}{Q^{SM}_{III-V}}=
\frac{2\left[(g^\nu_V)^2 +(g^\nu_A)^2\right]\left[(g^e_V)^2 +(g^e_A)^2\right]- \left[(g^{eSM}_V)^2 +(g^{eSM}_A)^2\right]}{\left[(g^{eSM}_V)^2 +(g^{eSM}_A)^2\right]}.
\end{equation}

From this figure it is clear that the relative correction reaches its maximum value between $1.5 \leq f \leq 2 \hspace{0.8mm}TeV$,
and is of the order of $3.5\%$ with respect to the standard model \cite{Dicus}, and decreases rapidly for large $f$. The curves also
demonstrate that the effect of the SLHM is not sensitive to $f$ in the range of $f\geq 6.5 \hspace{0.8mm} TeV$. This is generally
because, the extra contribution of the SLHM model to the relative correction is proportional to a factor of $\frac{1}{f^2}$.

In general, the relative correction is sensitive to the parameters $\theta$, $\phi$ and $f$ of the models considered. However,
there are other effects which may change the stellar energy loss rates, for example, the radiative corrections at one-loop level.

We conclude that the energy loss via $\nu + \bar\nu$ pairs is relevant at the moment of collapse when thermal
process become extremely important. Even when the stellar energy loss is dominated by heavy lepton flavor neutrinos, the
energy loss is higher in general for the threes extensions of the standard model, being maximum for the Simplest
Little Higgs Model (SLHM), up to $3.5\%$ in comparison with the SM. It is worth mentioning that at the present time
an  enhancement or suppression of individual weak rates on the order of $3.5\%$ cannot be identified. This means that it will leave
no imprint in neither the supernova dynamics nor the potentially observable neutrino signal. Therefore, some SM weak rates uncertainties
can be as large as one order of magnitude at some specific conditions, mainly due to the unknown state of matter of the supernova medium
in particular at high matter density. This can be improved considering that the analytical approximation for the Fermi integrals must be
performed for various limits regions of high matter density and then evaluating the relative correction $\frac{\delta Q}{Q^{SM}_{\nu\bar \nu}}= \frac{Q^{New}_{\nu\bar\nu} - Q^{SM}_{\nu\bar\nu}}{Q^{SM}_{\nu\bar\nu}}$.

In conclusion, in this article we determine exact and approximate analytical expressions for the stellar energy loss rates through
the process $e^+ e^- \to \nu \bar \nu$: in the context of a 331M, a LRSM and the SLHM. In addition, we study the contributions
of the parameters of these models through the relative correction and for different limiting cases as is mentioned in the text. We
find that the stellar energy loss rates is almost independent of the mixing angle $\theta$, $\phi$ and $f$ of each considered model
in the allowed range for these parameters. As expected, in the decoupling limit, when $\theta=0$, $\phi=0$  and
$f \to \infty$, the expression for the stellar energy loss rates $Q^{SM}_{\nu\bar\nu}\left(\beta\right)$ of the SM previously obtained
in the literature \cite{Armando,Yakovlev,Esposito1,Esposito2,Misiaszek} is recovered. Furthermore, our analytical and numerical results
for the stellar energy loss rates have never been reported in the literature before, and complements other studies on the stellar energy
loss rates in $e^+e^-$ annihilation and could be of useful for the scientific community.
In the calculation of the stellar energy loss rates was needed the computation of the Fermi integral in different
regions of density and temperature. These Fermi integrals, and its implementation in large-scale astrophysics simulations
as well as in the study of the stellar energy loss rates will be published in a paper in preparation \cite{Hernandez1}.

\vspace{1cm}

\begin{center}
{\bf Acknowledgments}
\end{center}

This work was supported by CONACyT, SNI and PROFOCIE (M\'exico). AGS thanks Lerma and Observatorire de Paris for a Visiting 
Astronomer position during which part of this work was carried on.


\newpage


\begin{figure}[t]
\centerline{\scalebox{0.65}{\includegraphics{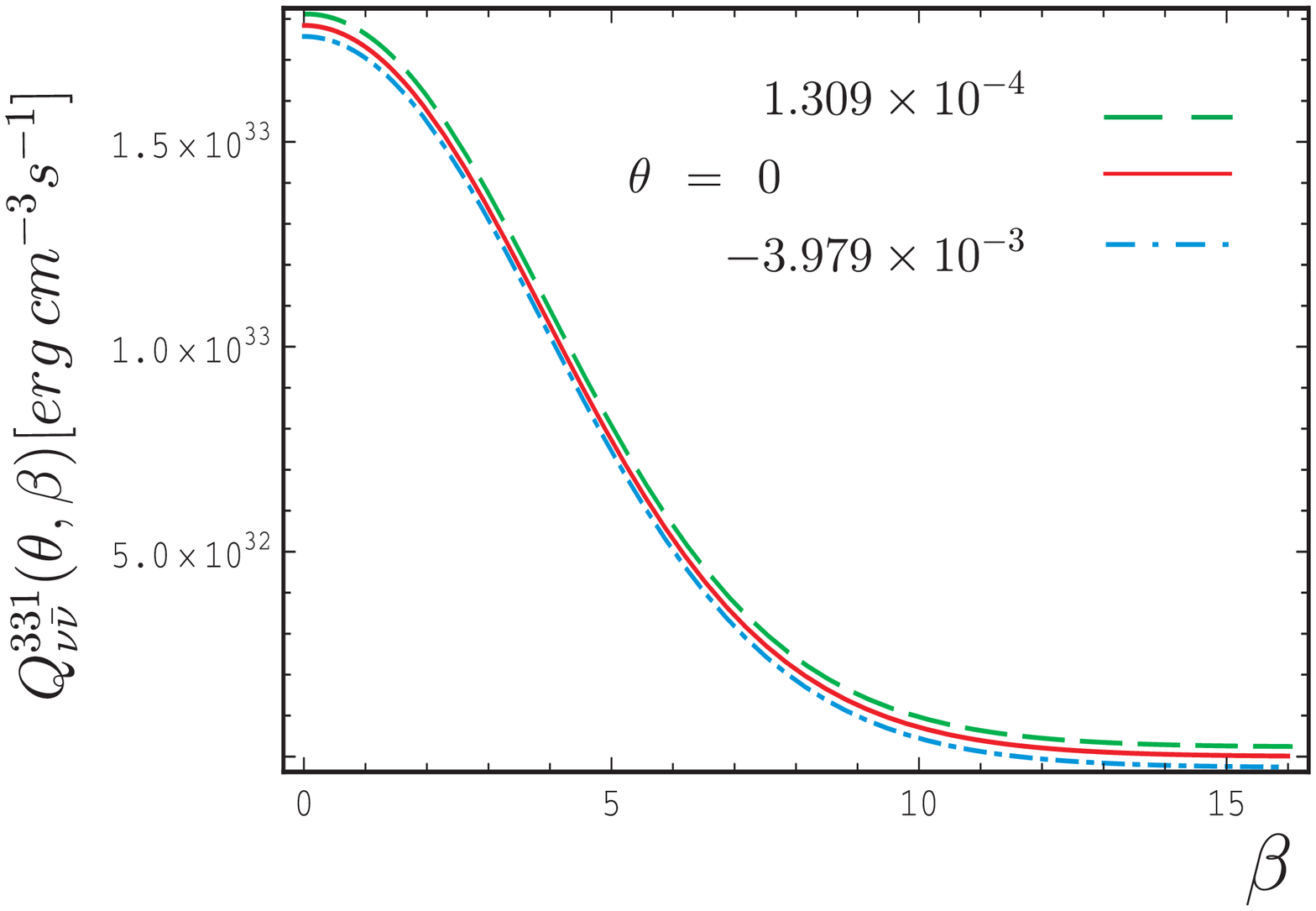}}}
\caption{ \label{fig:gamma} The stellar energy loss rates for $e^+ e^- \to \nu \bar \nu$ as a function of degeneration parameter $\beta$
and the mixing angle $\theta=-3.979\times 10^{-3},\hspace{0.8mm} 0,\hspace{0.8mm} 1.309\times 10^{-4}$ of the 331M.}
\end{figure}

\begin{figure}[t]
\centerline{\scalebox{0.65}{\includegraphics{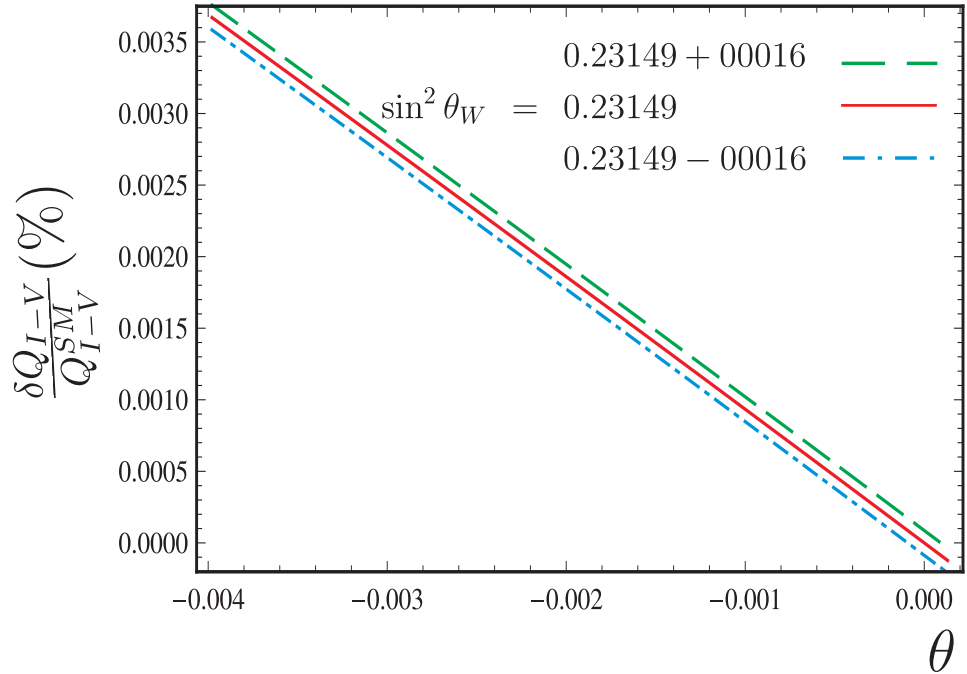}}}
\caption{ \label{fig:gamma} The relative correction
$\frac{\delta Q_{I-V}}{Q^{SM}_{I-V}}=
\frac{a^2b^2\left[(g^e_V)^2 +(g^e_A)^2\right]- \left[(g^e_V)^2 +(g^e_A)^2\right]}{\left[(g^e_V)^2 +(g^e_A)^2\right]}$
as a function of $\theta$ and $\sin^2\theta_W=0.23149-0.00016,\hspace{0.8mm} 0.23149,\hspace{0.8mm} 0.23149+0.00016$.}
\end{figure}

\begin{figure}[t]
\centerline{\scalebox{0.65}{\includegraphics{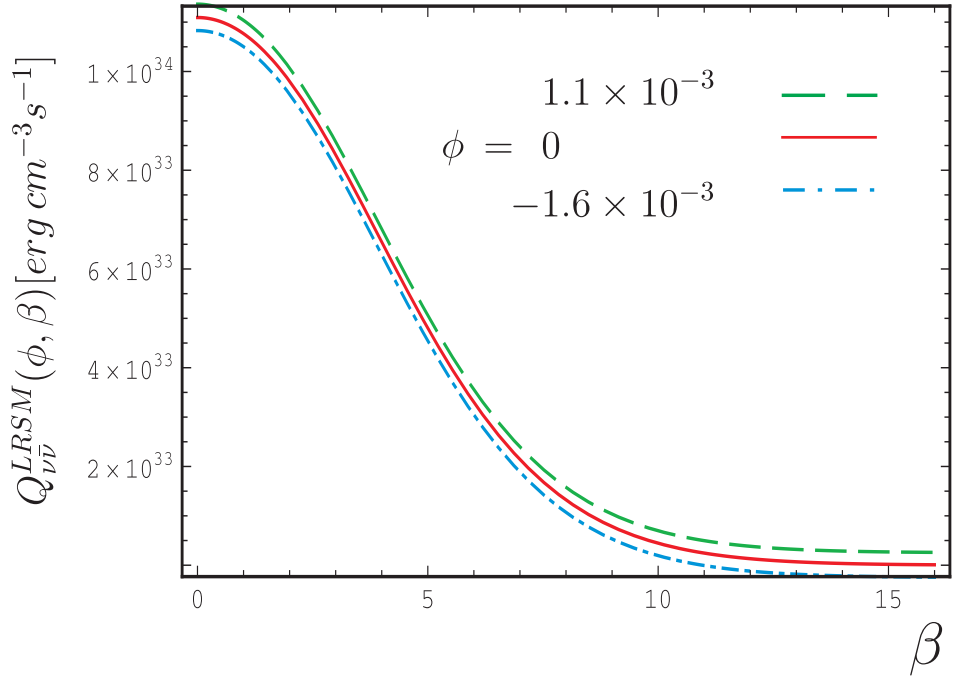}}}
\caption{ \label{fig:gamma} The same as Figure 1 but for $\beta$ and the mixing angle
$\phi=-1.6\times 10^{-3},\hspace{0.8mm} 0,\hspace{0.8mm} 1.1\times 10^{-3}$ of the LRSM.}
\end{figure}

\begin{figure}[t]
\centerline{\scalebox{0.65}{\includegraphics{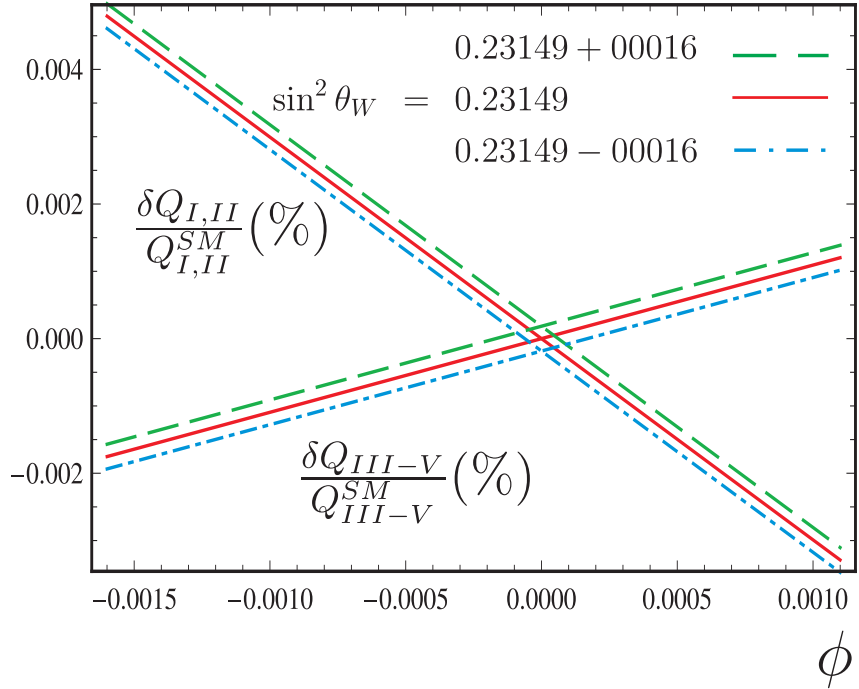}}}
\caption{ \label{fig:gamma} The same as Figure 2 but for
$\frac{\delta Q_{I-II}}{Q^{SM}_{I-II}}=
\frac{a'^2(a'^2 + b'^2)(g^e_V)^2 - 2(g^e_V)^2}{2(g^e_V)^2}$ and
$\frac{\delta Q_{III-V}}{Q^{SM}_{III-V}}=
\frac{(a'^2+b'^2)\left[a'^2(g^e_V)^2 +b'^2(g^e_A)^2\right]- 2\left[(g^e_V)^2 +(g^e_A)^2\right]}{2\left[(g^e_V)^2 +(g^e_A)^2\right]}$
as a function of $\phi$ and $\sin^2\theta_W=0.23149-0.00016,\hspace{0.8mm} 0.23149,\hspace{0.8mm} 0.23149+0.00016$.}
\end{figure}

\begin{figure}[t]
\centerline{\scalebox{0.65}{\includegraphics{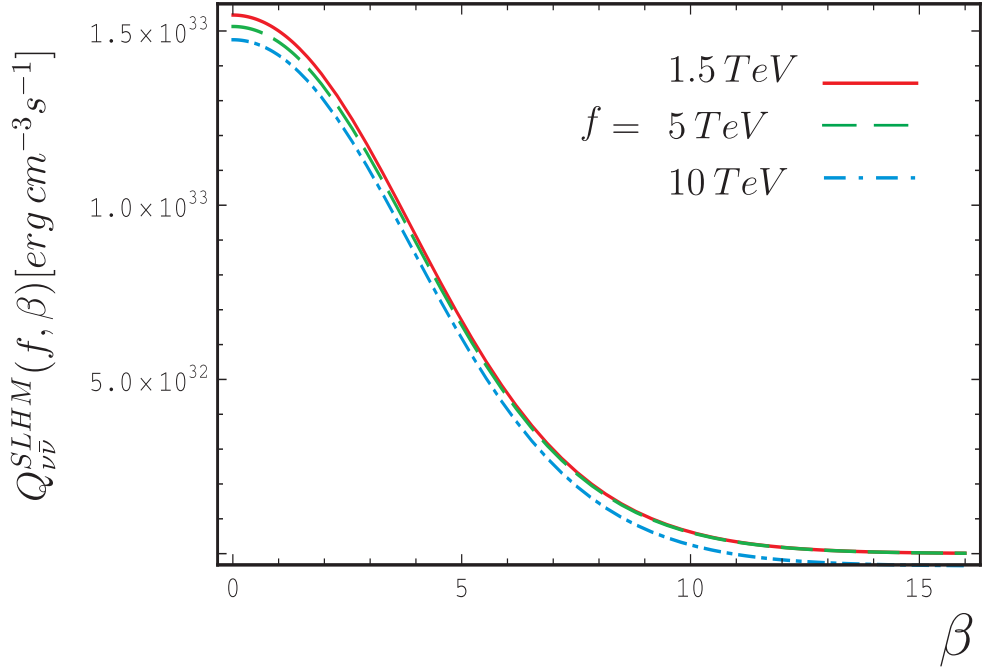}}}
\caption{ \label{fig:gamma} The same as Figure 1 but for $\beta$ and the characteristic energy scale $f= 1.5\hspace{0.8mm}TeV,\hspace{0.8mm} 5\hspace{0.8mm}TeV, 10\hspace{0.8mm}TeV$ of the SLHM.}
\end{figure}

\begin{figure}[t]
\centerline{\scalebox{0.65}{\includegraphics{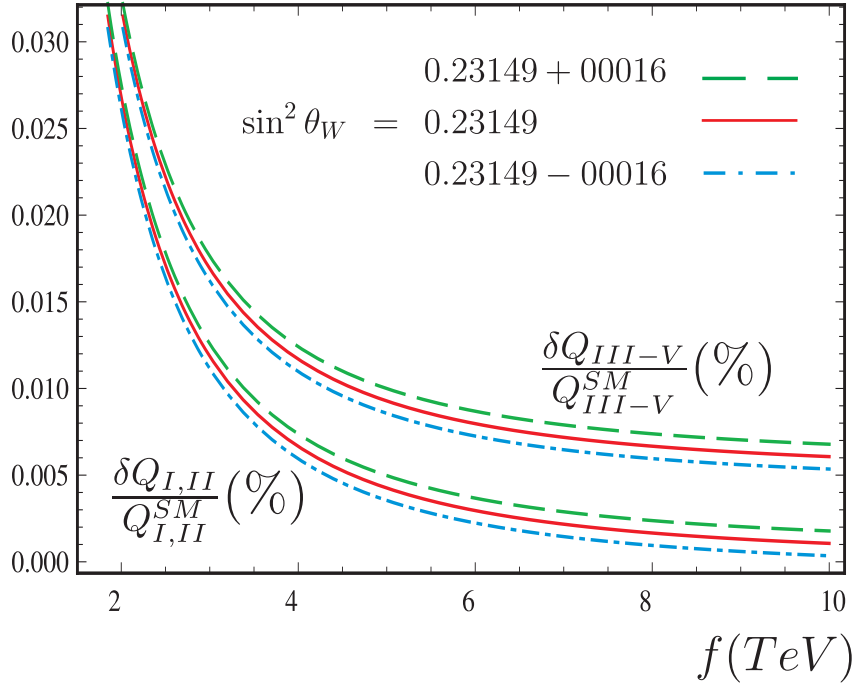}}}
\caption{ \label{fig:gamma} The same as Figure 2 but for
$\frac{\delta Q_{I-II}}{Q^{SM}_{I-II}}=
\frac{2(g^e_V)^2\left[(g^\nu_V)^2 +(g^\nu_A)^2\right]-(g^{eSM}_V)^2}{(g^{eSM}_V)^2}$ and
$\frac{\delta Q_{III-V}}{Q^{SM}_{III-V}}=
\frac{2\left[(g^\nu_V)^2 +(g^\nu_A)^2\right]\left[(g^e_V)^2 +(g^e_A)^2\right]- \left[(g^{eSM}_V)^2 +(g^{eSM}_A)^2\right]}{\left[(g^{eSM}_V)^2 +(g^{eSM}_A)^2\right]}$
as a function of $f$ and $\sin^2\theta_W=0.23149-0.00016,\hspace{0.8mm} 0.23149,\hspace{0.8mm} 0.23149+0.00016$.}
\end{figure}

\end{document}